%% file: main.tex
\documentclass[runningheads]{llncs}
\usepackage[T1]{fontenc}
\usepackage{graphicx}

\usepackage{algpseudocodex}
\usepackage{algorithm}

\usepackage{epsfig}
\usepackage{amsbsy}
\usepackage{subcaption}
\usepackage{url}
\usepackage{color,soul}
\usepackage{xcolor}

\usepackage{amsmath}
\usepackage{amsfonts}

\begin{document}

\input{sections/title}
\input{sections/abstract}
\input{sections/introduction}

\input{sections/methodology}

\input{sections/experiments}

\input{sections/discussion}

%
%
%
\bibliographystyle{splncs04}
\bibliography{citation}

\end{document}

%% file: sections/title.tex
\title{ReeSPOT: Reeb Graph Models Semantic Patterns of Normalcy in Human Trajectories}
%
%
\author{Bowen Zhang\thanks{Equal Contributors} \and
S. Shailja\protect\footnotemark[1] \and
Chandrakanth Gudavalli \and 
Connor Levenson \and 
Amil Khan \and
B. S. Manjunath
}

%

\authorrunning{Zhang et al.}
\titlerunning{ReeSPOT}

%
\institute{Electrical and Computer Engineering Department \\ University of California Santa Barbara \\
 }
\maketitle              

%% file: sections/abstract.tex
\begin{abstract}

This paper introduces ReeSPOT, a novel Reeb graph-based method to model patterns of life in human trajectories (akin to a fingerprint). Human behavior typically follows a pattern of normalcy in day-to-day activities. This is marked by recurring activities within specific time periods. In this paper, we model this behavior using Reeb graphs where any deviation from usual day-to-day activities is encoded as nodes in the Reeb graph. The complexity of the proposed algorithm is linear with respect to the number of time points in a given trajectory. We demonstrate the usage of ReeSPOT and how it captures the critically significant spatial and temporal deviations using the nodes of the Reeb graph. Our case study presented in this paper includes realistic human movement scenarios: visiting uncommon locations, taking odd routes at infrequent times, uncommon time visits, and uncommon stay durations. We analyze the Reeb graph to interpret the topological structure of the GPS trajectories. Potential applications of ReeSPOT include urban planning, security surveillance, and behavioral research.  


\keywords{Reeb Graphs \and Graph Networks \and Trajectory Analysis}
\end{abstract}

%% file: sections/introduction.tex
\section{Introduction}
\label{sec:intro}

\begin{figure}[t]
  \centering
  \includegraphics[width=0.9\textwidth]{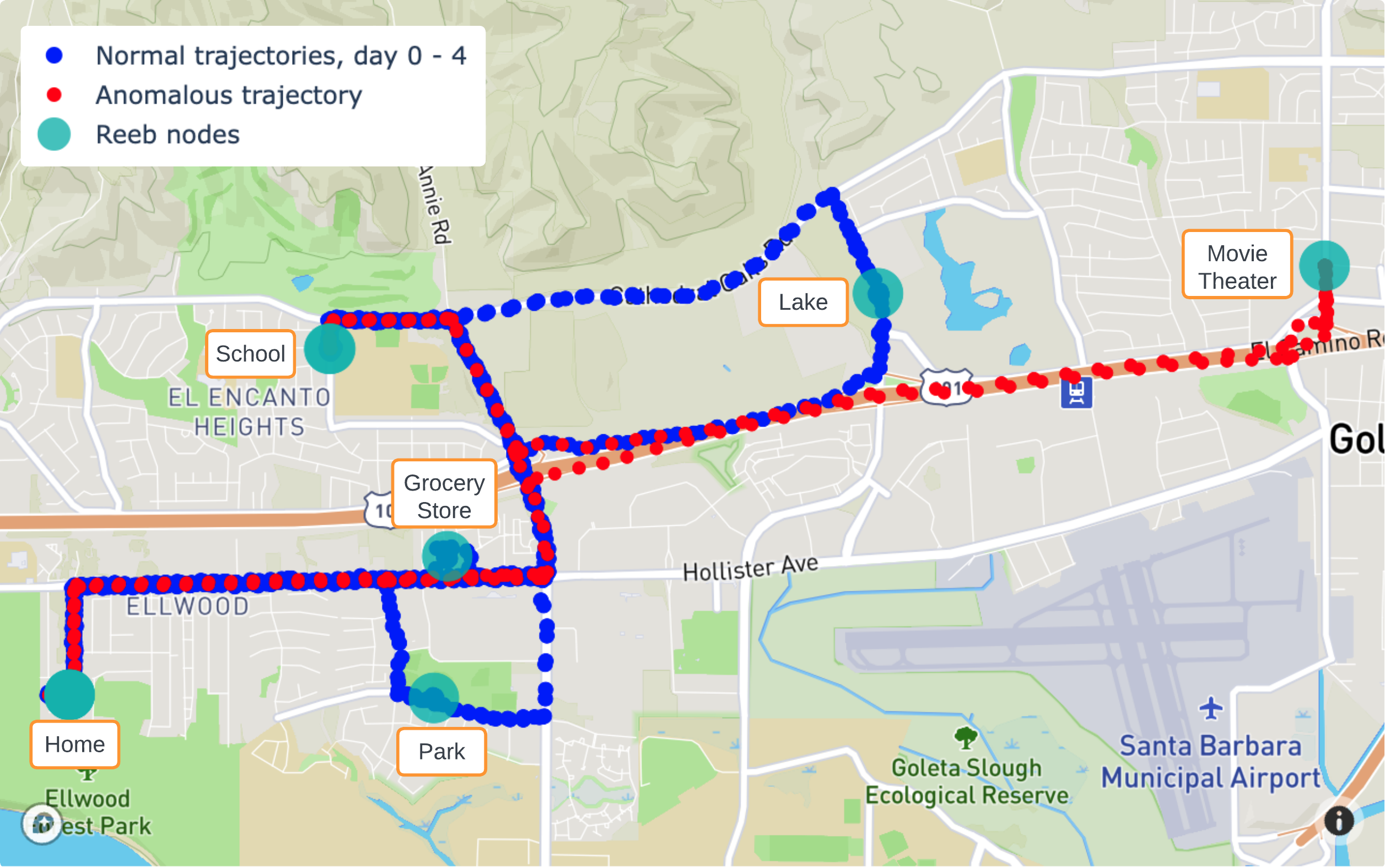}
  \caption{Map overlay of normal and anomalous trajectories from scenario 2 of the case study, annotated with semantic labels for points of interest (POIs).}
  \label{fig:map}
\end{figure}

Recently, there has been an increase in location-aware devices that use the Global Positioning System (GPS) for many applications such as finding efficient routes~\cite{ta2016built}, fitness apps, understanding the progression of infectious diseases~\cite{hast2019use}, and predicting demographic information~\cite{wu2019inferring}. This collection of movements, and thus vast amounts of raw trajectories, spotlights the need for a scalable representation of these trajectories that preserves and highlights the structure and topologically important movement patterns (Figure~\ref{fig:map}).

Human movement analysis is the core component of behavioral research, urban planning, and computational sociology~\cite{eagle2006reality}, which helps in better modeling human behavior and predicting human movement patterns. Similarly, modeling normal human behavior can also help identify abnormal human behavior. 
In particular, given a set of movement patterns for a week, month, or year, we want to capture any change in the semantic ``patterns of life''. In this paper, we model routine behaviors and movements that characterize daily human activities in a given city using a concept from topology, Reeb graphs.


Traditional trajectory analysis methods are largely based on hand-crafted geometric features and statistical techniques. Such features include traveling distance, mean velocity~\cite{zheng2008learning}, frequencies of areas or moving patterns~\cite{giannotti2007trajectory}. Statistical approaches analyze the temporal patterns with respect to the frequency of trajectory data to identify patterns such as traveling modes~\cite{kinoshita2016gps} and periodic patterns~\cite{zhang2019mining}. These approaches are effective for handling structured and less complex data sets but fail to generalize with high-dimensional data or the dynamic nature of human mobility patterns.

Given the amount of GPS data that can be generated by one human on a single day, another obvious direction to look at would be toward data-driven learning methods. Specifically, sampling a single agent's movement data, sampled at a 1Hz frequency over a month, accumulates roughly 2 million data points. 

Extrapolating these figures to a population of a small city like Santa Barbara, with approximately 97,000 agents, results in a dataset comprising an immense 194 billion data points. This scale poses substantial challenges in terms of computational resources and data management, and extrapolating to larger cities, such as New York City, would significantly magnify these challenges.
Recent advances in deep learning have significantly enhanced the capability to model human mobility patterns by performing the next-location prediction~\cite{luca2021survey}. Particularly, long short-term memory networks (LSTMs) \cite{hochreiter1997long} and attention-based models like Transformers \cite{vaswani2017attention} are good at capturing temporal regularities and anomalies in movement patterns. However, these black-box models lack interpretability, thus limiting their applicability in real-time scenarios~\cite{zeng2019next}. 


Towards interoperability along with large-scale modeling, Graph-based methods are very popular due to their ability to represent complex spatial relationships and movement patterns efficiently. We need models that can succinctly summarize an agent’s trajectory data---retaining essential information while discarding redundancies. Transforming GPS data into graph data structures with nodes as significant geographic locations and edges as the movement information between enables intuitive models for pattern-of-life. Research directions include, Guo et al. \cite{guo2010graph}'s graph model to establish precise topological relationships among trajectories and geographic locations. Qi et al. \cite{qi2015efficient} incorporate hybrid methods that blend graph-based approaches with statistical models to improve the accuracy of trajectory searches and predictions. Another such work focuses on hierarchical clustering based on graph similarity measures\cite{sabarish2020graph}, further supporting the need for computational geometry. 

In this paper, we use Reeb graphs to cluster the common behavior pattern for a given agent. Our research is motivated by and related to previous research on the construction of Reeb graphs for trajectory data~\cite{buchin2013trajectory,shailja2023reebundle}. A Reeb graph captures the connectivity of level sets of a scalar function defined over a space, effectively summarizing the topological features of the space. In the context of trajectory data, scalar functions could represent attributes such as speed, direction, semantics, or geographical points of interest. Reeb graphs can thus map complex trajectories into more interpretable topological constructs. This abstraction facilitates the detection of anomalies by comparing the topological signatures of trajectories and identifying those that differ significantly from the norm. Our main contributions are summarized below:
\begin{itemize}
    \item We propose a novel Reeb graph-based approach to model the day-to-day activities of a given agent. To the best of our knowledge, this is the first demonstration of Reeb graphs to fingerprint an agent's behavior.
    \item We discuss the algorithm and its time complexity demonstrating the scalability of the proposed method.
    \item We design normal and anomalous scenarios, describe the methods for trajectory generation and present detailed experiments on the interpretation and analysis of Reeb graphs.
\end{itemize}

%% file: sections/methodology.tex
\section{Methodology}
\label{sec:prop_method}

\subsection{Previous work on Reeb graphs}
Reeb graph was first proposed to study the topology of a manifold~\cite{shinagawa1991surface}. Nodes of the Reeb graph encode the evolution of the level sets of a real-valued function on a manifold. The location of the node is the average location of the points of the trajectories that constitute the node. Reeb graphs have been extensively used in shape analysis for diverse datasets~\cite{biasotti2008reeb}. The first study of Reeb graphs for trajectory group evolvement encodes the merging and splitting structure between different moving entities~\cite{buchin2013trajectory}. Similarly, the spatial subtrajectory clustering algorithm presented a stricter problem~\cite{shailja2023reebundle,shailja2023retrace,shailja2021computational} but discovers geometric and topological substructure. This is a computationally challenging problem because the initialization step involves an exhaustive search of an agent's events. Motivated by these challenges, the central focus of this paper is to develop a method for fingerprinting the behavior of an agent over time such as days, weeks, and months. Our approach encodes significant spatio-temporal points of interest—specifically, locations and durations that define critical aspects of an agent's behavior. We redefine the grouping definitions used in our adapted Reeb graph model. The constructed Reeb graphs effectively partition a set of GPS points into meaningful nodes and edges, thereby quantifying and identifying path deviations. 

\begin{algorithm}[b!] 
\caption{Find \textit{connect} and \textit{disconnect} events}
\label{alg:event_compute}
\begin{algorithmic}[1]
\State \textbf{Input:} Trajectories $T$ and $T'$, threshold $\epsilon$
\State \textbf{Output:} Dictionary of \textit{connect}/\textit{disconnect} events, $events_{T,T'}$

\State Initialize $events_{T,T'}$ as an empty dictionary
\State Initialize $k \gets 0$
\State Initialize $connect\_flag \gets \text{False}$
\While{$k < m$}
    \If{$d(T[t_k], T'[t_k]) < \epsilon$}
        \State $events_{T,T'}[t_k] \gets \text{\textit{connect}}$
        \State $connect\_flag \gets \text{True}$
        \While{$k < m$ and $d(T[t_k], T'[t_k]) < \epsilon$}
            \State $k \gets k + 1$
        \EndWhile
        \If{$k < m$}
            \State $events_{T,T'}[t_k] \gets \text{\textit{disconnect}}$
        \EndIf
    \EndIf
    \State $k \gets k + 1$
\EndWhile
\State \Return $events_{T,T'}$
\end{algorithmic}
\end{algorithm}

\subsection{Reeb graph models agent pattern of normalcy}
A trajectory $T$ is defined as a dictionary (key: value) containing an ordered sequence of time points and their associated GPS coordinates: 
\begin{align}
    T = \{t_0 : p_0, t_1 : p_1, t_2 : p_2, \ldots, t_m : p_m\},
\end{align}
where $m$ is chosen according to the desired resolution to sample the pattern of the agent. Here  \( m \) denotes the total number of points in a given trajectory \( T \). The frequency of GPS data sampling decides \( m \). For example, to model the weekdays of an agent's activities, the raw GPS data is sampled every second, giving us  \( m = 86400  \) which is the total number of seconds in a day. Similarly, if the data is sampled every hour, then \( m = 24 \) points per day. We define $n$ as the total number of trajectories for a given agent. For example, to model month-long data, $n = 30$ and for weekdays, $n = 5$. The common setting used throughout the paper for our problem definition is \( m = 24 \) and \( n = 5 \). Each time point $t_i$ corresponds to a GPS coordinate $p_i$ representing the position of the agent at time $t_i$. $p_i = (\text{lat}_i, \text{lon}_i)$, where $\text{lat}_i$ represents the latitude and $\text{long}_i$ represents the longitude. The Euclidean distance between two GPS coordinates $p_i$ and $p_{i'}$ is calculated at time $t_i$ as follows:
\begin{align}
    d(p_i, p'_{i}) = \sqrt{(\text{lat}_i - \text{lat}'_{i})^2 + (\text{lon}_i - \text{lon}'_{i})^2},
\end{align}
where $\text{lat}_i$ and $\text{lon}_i$ are the latitude and longitude of the first point, and $\text{lat}'_{i}$ and $\text{lon}'_{i}$ are those of the second point. $d(p_i, p'_{i}) $ gives the 2-norm distance between two points on the Euclidean plane. This approximates the geographic distance of the points. The algorithm is defined with respect to a distance threshold $\epsilon$ within which the points are considered sufficiently close together i.e. within a small geographical area. This is the inter-trajectory distance that guides the granularity of the Reeb graphs according to the problem definition.
\begin{figure}[!h]
  \centering
  \includegraphics[width=0.9\textwidth]{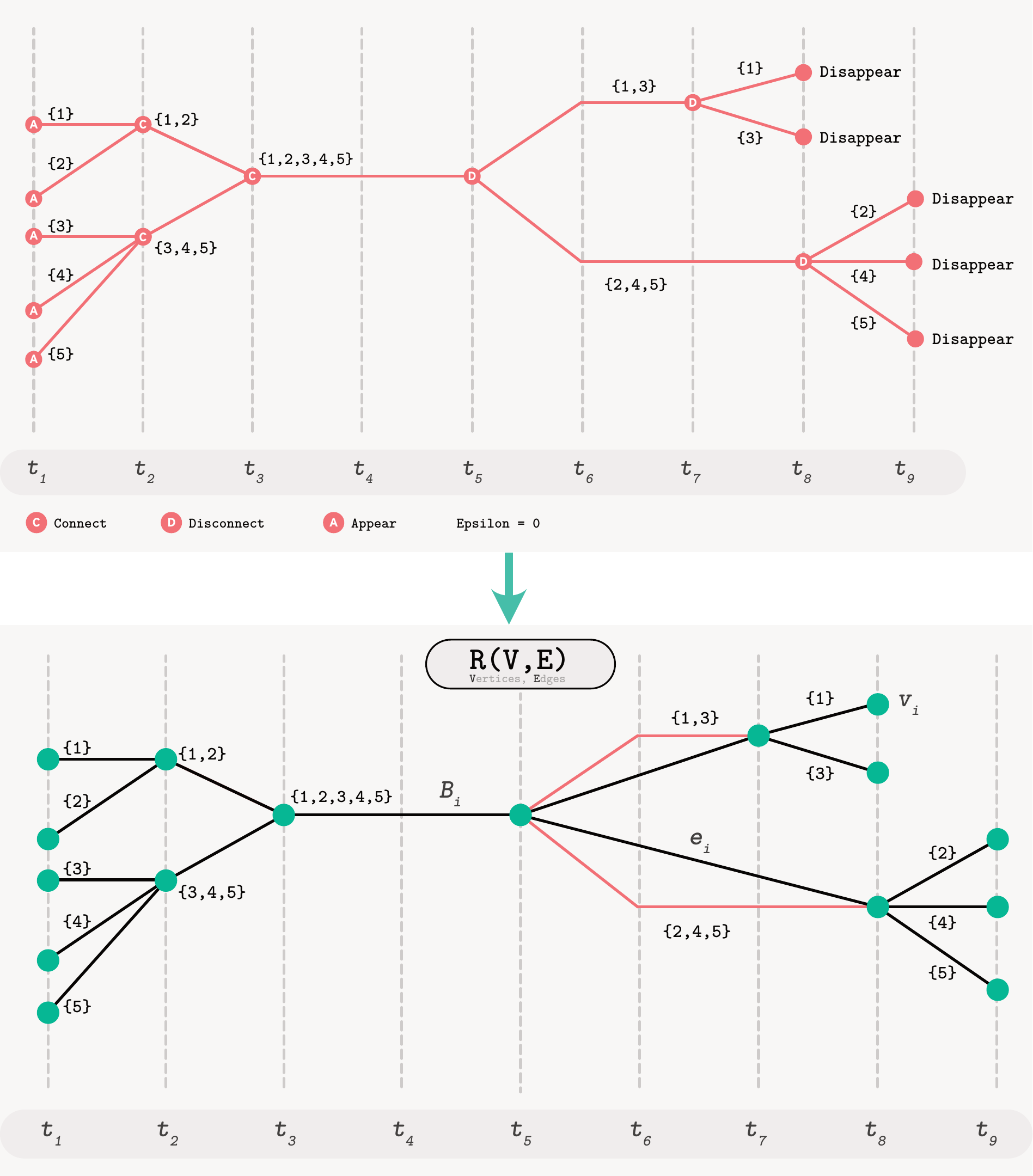}
  \caption{\textsc{Reeb Graph Construction Over Time.} We show the construction of Reeb graphs $R(V,E)$ for a set of five trajectories. The \textit{appear, disappear, connect,} and \textit{disconnect} events are shown in the top subfigure. Changes in the grouping of trajectories due to these events are encoded as nodes in the bottom figure. Nodes of the Reeb graph $\mathcal{R}$ in the bottom figure are shown in green color and the edges are shown in black color.
  }
  \label{fig:reeb}
\end{figure}

Human behavior typically follows a pattern of normalcy in day-to-day activities. This is marked by recurring activities within specific time periods. In order to discover the large-scale spatio-temporal patterns, we represent the bundling structure of trajectories as a \textit{Reeb graph} $R(V,E)$. Nodes of the Reeb graph will pinpoint critical GPS points of the agent's pattern. Intuitively, if a continuous portion of a behavior of the agent happens at the same time and within the same spatial distance ($\epsilon$) every day then they present a pattern of normalcy. We formalize this by introducing the concept of ``bundles'' to characterize normal behavior through consistent daily subtrajectory events. Each trajectory begins with an \textit{appear} event at the first index and concludes with a \textit{disappear} event at the last index of $T$. Deviations from this norm by more than $\epsilon$ are classified as \textit{disconnect} events, while a return to the norm is labeled a \textit{connect} event. Formally, for a given $\epsilon$ and $m = 23$ i.e. sampled every hour, let's take two trajectories $T$ and $T'$:
\begin{itemize}
    \item At time $t_0$: $p_0$ and $p'_0$ are the \textit{appear} events.
    \item At time $t_{23}$: $p_{23}$ and $p'_{23}$ are the \textit{disappear} events.
    \item If $d(p_0, p'_0) \leq \epsilon, (p_1, p'_1) \leq \epsilon, \ldots, d(p_k, p'_k) \leq \epsilon$, but $d(p_{k+1}, p'_{k+1}) > \epsilon$, then $t_{k+1}$ represents a \textit{disconnect} event between $T$ and $T'$.
    \item If $d(p_0, p'_0) > \epsilon, (p_1, p'_1) > \epsilon, \ldots, d(p_k, p'_k) > \epsilon$, but $d(p_{k+1}, p'_{k+1}) \leq \epsilon$, then $t_{k+1}$ represents a \textit{connect} event between $T$ and $T'$.
\end{itemize}

\begin{algorithm}[t!]

\caption{Construction of Reeb Graph}
\label{alg:event_handling}
\begin{algorithmic}
 \Function{ConstructReebGraph}{set of events for all pairs of trajectories ($E$)}
\ForAll{steps $k$ from 0 to |E|} \Comment{Dynamic Graphs}
\If{appear event of $T$}
\State insert new node $T$ to $G_{k}$
\EndIf
\If{disappear event of $T$}
\State delete node $T$ from $G_{k}$
\EndIf
\If{connect event between $T_x$ and $T_y$}
\State insert edge $(T_x, T_y)$ to $G_{k}$
\EndIf
\If{disconnect event of trajectories $T_x$ and $T_y$}
\State delete edge $(T_x, T_y)$ from $G_{k}$
\EndIf
\State $P \gets$ empty bundle partition\Comment{Bundle Partition}
\State Query $G_{k-1}$ and $G_{k}$ to get the connected components $C_{k-1}$ and $C_{k}$ respectively;

\ForAll{connected component $c_k \in C_k$}
\If{$c_k \in C_{k-1}$}
\State assign the same bundle id $B_i$ to the points for trajectories in $c_k$;
\Else
 \State create a new bundle id $B_{i+1}$ and assign it to the points for trajectories in $c_k$;
\EndIf
\State Add $B_{i+1}$ to $P$
\EndFor
\EndFor
\State Construct Reeb graph $R$ from $P$ by connecting adjacent bundles with nodes and bundles as edges; \Comment{Construct Reeb graph}

\Return $R$
\EndFunction
\end{algorithmic}

\end{algorithm}

\subsection{Construction of Reeb graphs and analysis of time complexity}

Reeb graph construction (illustrated in Figure~\ref{fig:reeb}) can be divided into the following major steps: event computation, construction of dynamic graphs ($G$s), connectivity query in the dynamic graph for bundle partition ($P$), and construction of the Reeb graphs ($R$) from bundles partition as shown in Figure~\ref{fig:reeb}. The first step of Reeb graph construction involves computing the \textit{connect} and \textit{disconnect} events. Algorithm~\ref{alg:event_compute} outlines the steps of computing events. The event computation takes $\mathcal{O}(m)$ time, where \(m\) represents the number of time points in the trajectories \(T\) and \(T'\). At each time point, the algorithm looks for $\mathcal{O}(5\times 5)$ possibilities of potential events. The second step of the Reeb graph involves handling the events to construct dynamic graph $G$s. The nodes of $G$ represent the daily trajectories and the edges of the $G$ represent the $\epsilon$-connectivity between them. The total number of nodes in $G$ is 5 representing one trajectory for each day of the agent. The connected component of the $G$ will give us the $\epsilon-$step bundle partition of subtrajectories denoted by \(P = \{B1, B2, \ldots, Bk\}\) such that every segment in ${T_0, T_1, T_2, T_3, T_4}$ is uniquely assigned to exactly one bundle. The final step is to construct the Reeb graph from these bundles. Reeb graph $R$ can be constructed from $P$ by connecting adjacent bundles with nodes and bundles as edges similar to the described construction in ~\cite{shailja2023reebundle}. So, the time complexity of the Reeb graph construction step would be $\mathcal{O}(m)$ because in the worst case, all the time points will have events. At each time, the connectivity query to the dynamic graph with 5 nodes takes constant time. The more detailed steps can be found in the Algorithm~\ref{alg:event_handling}.






%% file: sections/experiments.tex
\section{Experimentation/Case Study}
\label{sec:experiments}

\subsection{Data generation}
We model the pattern of life of a single agent over different trajectories. Each trajectory is simulated using the SUMO software package \cite{SUMO2018} and represents realistic behavior and movement patterns over the course of one week. We construct the Reeb graph for each trajectory and show how it sufficiently represents the trajectory's information with significantly fewer nodes. 
\begin{figure}[h]
  \centering
  \includegraphics[width=0.85\textwidth]{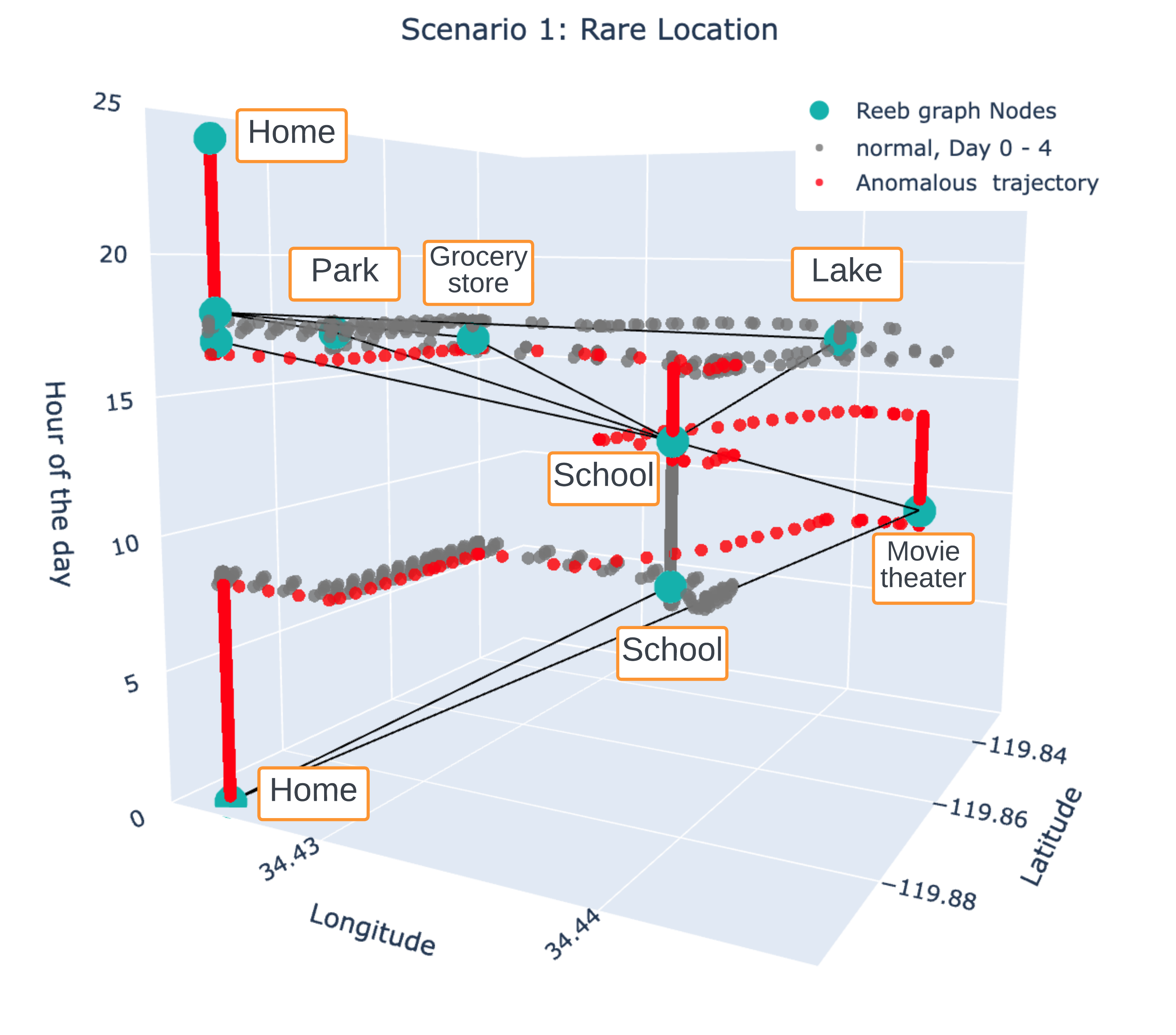}
  \caption{3D trajectory plots with computed Reeb graph nodes for scenario 1 in Section \ref{sec:experiments}, where day 0 to day 4 are normal trajectories, and the anomalous trajectory is in red.}
  \label{fig:scenario1_3d}
\end{figure}

In this case study, we analyze the behavioral patterns of a simulated high-school student from the city of Santa Barbara, California (Figure \ref{fig:map}), using trajectory data that includes multiple points of interest (POIs), such as the student's home, school, park, grocery store, and lake. The student's daily routine typically consists of attending school from approximately 8:00 AM to 9:00 AM, concluding at around 4:00 PM to 5:00 PM, followed by visits to recreational sites before returning home. To thoroughly investigate both normal and anomalous behavioral patterns, we generated five days of normal trajectory data, complemented by additional days tailored to each specific scenario described earlier. Each trajectory entry is recorded with timestamps, latitude, and longitude coordinates. Figure~\ref{fig:map} displays the student's trajectories across different POI locations for the rare location scenario, illustrating the distribution of both routine and deviant movements. Figure~\ref{fig:scenario1_3d} displays the same data as a 3D plot, providing a clear spatio-temporal visualization of the student's stay locations, duration, and revisit frequencies.

\subsection{Definition of anomalous behavior}
\label{ssec:scenario}
We define $L$ as a set of normal POIs and their corresponding time points, \[
L = \{(lat_1, lon_1, t_1), (lat_2, lon_2, t_2), \dots, (lat_n, lon_n, t_n)\}
\]
where \( (lat_i, lon_i) \) represents the geographic coordinates with \( lat_i \in [-90, 90] \) and \( lon_i \in [-180, 180] \), and \( t_i \) is the time at which these coordinates were recorded. Relative to this definition, all the anomaly behaviors for a given agent are defined as follows:
\begin{algorithm}[t!]
\caption{Trajectory Generation}
\begin{algorithmic}[1]
\State \textbf{Inputs:}
\State \quad POIs -- List of Points of Interest as coordinates on a map.
\State \quad $TimeList_n$ -- Dictionary mapping each POI to normal visit times.
\State \quad $TimeList_a$ -- Dictionary mapping each POI to abnormal visit times.
\State \quad Road Network -- Road network graph for route generation.
\State \textbf{Output:}
\State \quad $T$ -- A list of normal trajectories of an agent visiting specified POIs.
\State \quad $T^*$ -- A list of abnormal trajectories of an agent visiting specified POIs.
\State Initialize Trajectories list
\For{each POI in the POIs list}
    \State Select $TimeList$ based on a decision rule (normal vs abnormal)
    \For{each time in $TimeList$}
        \State Generate a starting point for the agent
        \State Use duarouter to calculate the shortest path from the starting point to the POI at the given time
        \State Pass the list of edges to SUMO for movement simulation
        \State Collect the output trajectory from SUMO
        \State Append to $T$ or $T^*$ based on decision rule
    \EndFor
\EndFor
\State \Return $T$, $T^*$ 
\end{algorithmic}
\end{algorithm}

\textbf{{Scenario 1 (S1): Rare Location Anomaly}} Rare location anomaly refers to a scenario when an agent visits a new location $( lat^*, lon^*, t_i) \notin L$.  $( lat^*, lon^*)$ is spatially different from their normal spatial geographical points of interest such as school or work. Reeb graph will encode this rare location by creating a new node localizing the abnormality. 

\textbf{Scenario 2 (S2): Rare Route Visit Anomaly} In this scenario, the agent visits the same POI locations multiple times but utilizes a uniquely different route on a single journey. This introduces \textit{disconnect} event from their normal movement pattern, resulting in a new node in the Reeb graph. More formally, if $(lat^*, lon^*, t_{k:l}) \notin (lat, lon, t_{1:k-1}) $ and $\notin (lat, lon, t_{l+1:m}) $, then nodes $v_k$ and $v_l$ will be added to $R$.

\textbf{Scenario 3 (S3): Uncommon Time Visit} This is a case of time violation where the agent visits a familiar location at an uncommon time $t^*$ i.e, $( lat_i, lon_i, t^*) \neq ( lat_i, lon_i, t_i) $

\textbf{Scenario 4 (S4): Uncommon Stay Duration Anomaly} In this scenario the agent stays for an abnormal duration ($\Delta$) at a specific location $(lat^*, lon^*, t_{i+\Delta})$. This results in a \textit{disconnect} event for the agent's trajectory from the normal pattern of life at $t_i$. 


\subsection{Reeb Graph Generation}
We use a down-sampling rate of one hour for Reeb graphs. This setting helps us to monitor changes in location grouping states at each hour. The threshold $\epsilon$ for spatial connect and disconnect events is set to 0.0005 GPS degrees (5.56 meters). Initially, we construct a Reeb graph from the normal activity trajectories of days 0 to 4 to model the student’s typical pattern of life. 

As depicted in Figure~\ref{fig:map} and Figure~\ref{fig:scenario1_3d}, ReeSPOT successfully identifies all normal POIs as a part of the Reeb graph nodes, demonstrating its efficacy in reflecting the spatial distribution of the student’s activities. Notably, an anomalous scenario depicted in Figure~\ref{fig:map} and Figure~\ref{fig:scenario1_3d} shows the student visiting a movie theater during school hours which is defined as a deviation from the normal. This is captured by a new Reeb graph node, highlighting its potential for identifying critical spatial anomalies.
\begin{figure}[!ht]
  \centering
  \includegraphics[width=1\textwidth]{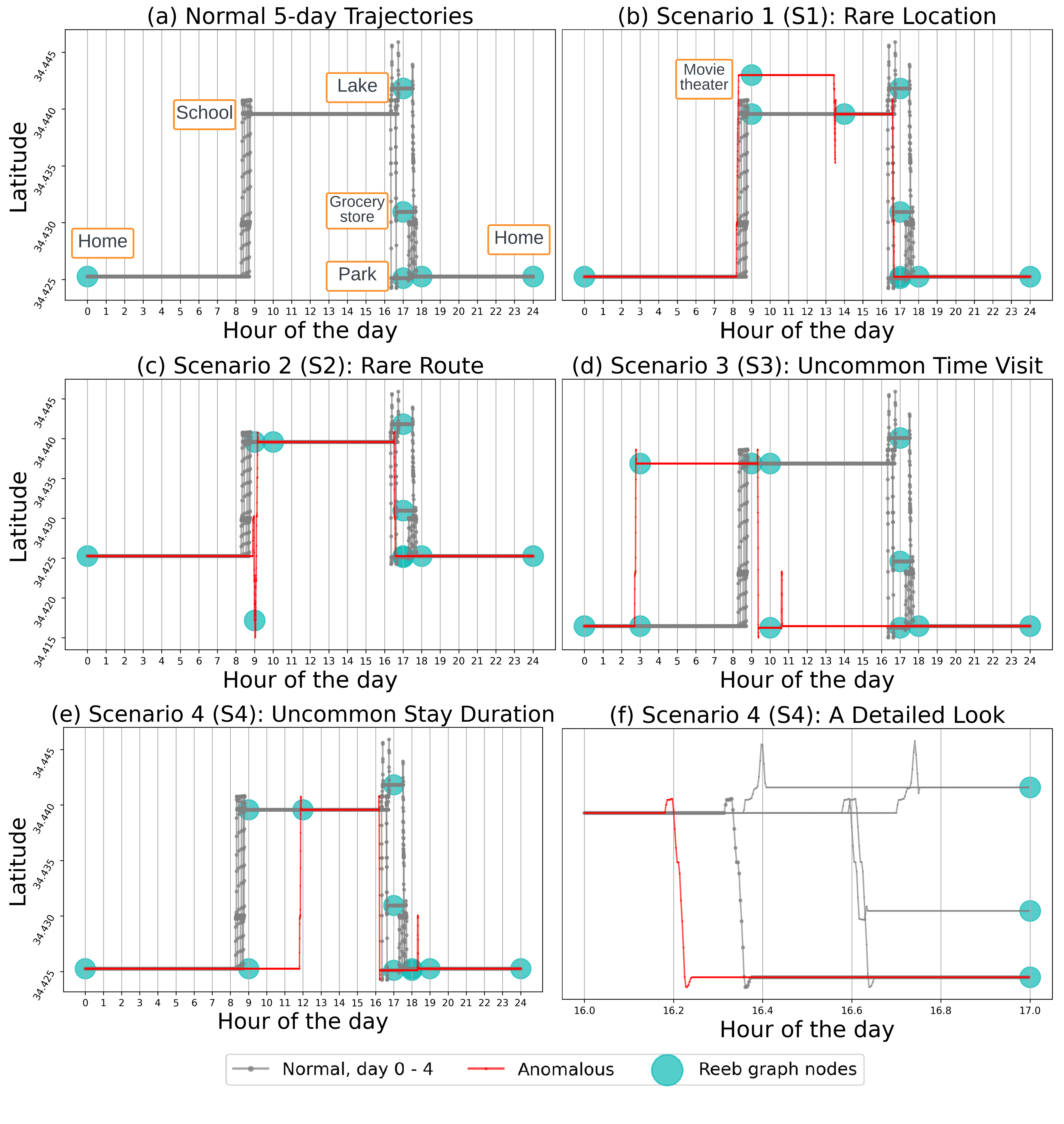}
  \caption{2D Trajectory plots displaying time and latitude dimensions alongside computed Reeb graph nodes. These plots illustrate both normal and anomalous scenarios as outlined in Section \ref{ssec:scenario}. The detailed discussions on node generation and behavioral analysis can be found in Section~\ref{ssec:plots_desc}.}
  \label{fig:scenarios_2d}
\end{figure}
\subsection{Analysis and interpretation of scenarios using Reeb graphs}
\label{ssec:plots_desc}
To better understand the formation of Reeb graph nodes and demonstrate the utility of the Reeb graph across all six scenarios, we generated time-latitude plots (Figure \ref{fig:scenarios_2d}). These plots, with the hour of day on the x-axis and latitude on the y-axis, include trajectory points sampled every 10 seconds alongside Reeb graph nodes. Each plot provides a visual representation of different behavioral patterns and anomalies and illustrates ReeSPOT's effectiveness in capturing anomalous trajectories for all scenarios. We explain the scenarios one by one below:


\begin{itemize}
    \item \textbf{Figure \ref{fig:scenarios_2d}(a)} illustrates the student’s normal routine pattern, with stays at home, school, and visits to various recreational spots. Notable events include \textit{appear} and \textit{disappear} at the beginning and end of each day. There are three \textit{disconnect} events around hour 17 which indicates divergences to different locations after school.\textit{Connect} event shows trajectories getting merged back on the way home at hour 18.
    \item \textbf{Figure \ref{fig:scenarios_2d}(b) for S1} depicts a rare location $(lat^*, lon^*)$ where we visualize an abnormal visit to the movie theater, showing three additional Reeb nodes and altered connectivity events at hour 9 and 14.
    \item \textbf{Figure \ref{fig:scenarios_2d}(c) for S2} captures an alternative route to school. At hour 9, instead of following the normal route, the student deviates towards a direction with a lower latitude and then returns to school. This deviation is captured by the bottom Reeb graph node at hour 9. Additionally, a \textit{disconnect} event occurs at 9, followed by a \textit{connect} event at hour 10 when all trajectories converge at the school. 

    \item \textbf{Figure \ref{fig:scenarios_2d}(d) for S3} reveals an uncommon time anomaly, where the student attends school at hour 2 and travels to the park at around hour 10, significantly deviating from the typical schedule, but with the same POIs.
    \item \textbf{Figure \ref{fig:scenarios_2d}(e) for S4} shows another time-related anomaly with a prolonged stay at home until almost hour 12, and similarly, 3 new nodes appear for the reeb graph because of \textit{disconnect} event from the usual trajectory. 
    \item \textbf{Figure \ref{fig:scenarios_2d}(f) for S4} presents a detailed look at scenario 4, from hour 16 to hour 17. Since the reeb graph sample rate is one hour, the reeb graph nodes appear at hour 17 to represent the \textit{disconnect} events in the past hour. 
\end{itemize}

\subsection{Reeb graph iteratively detects anomalous behavior of an agent}
In the context of detecting anomalous trajectories within real-life data (test dataset), we iteratively construct Reeb graphs on the test dataset to identify daily anomalous trajectories. An initial Reeb graph is constructed using training data with all normal trajectories. Subsequently, for each daily trajectory in the test dataset, the Reeb graph is iteratively updated day by day. To detect anomalous behaviors effectively, we compute the distance between the existing Reeb graph and every updated version that includes the additional daily trajectory. The subsequent section details our methodology for calculating this distance and presents the results derived from our case study.
\begin{figure}[t!]
  \centering
  \includegraphics[width=0.8\textwidth]{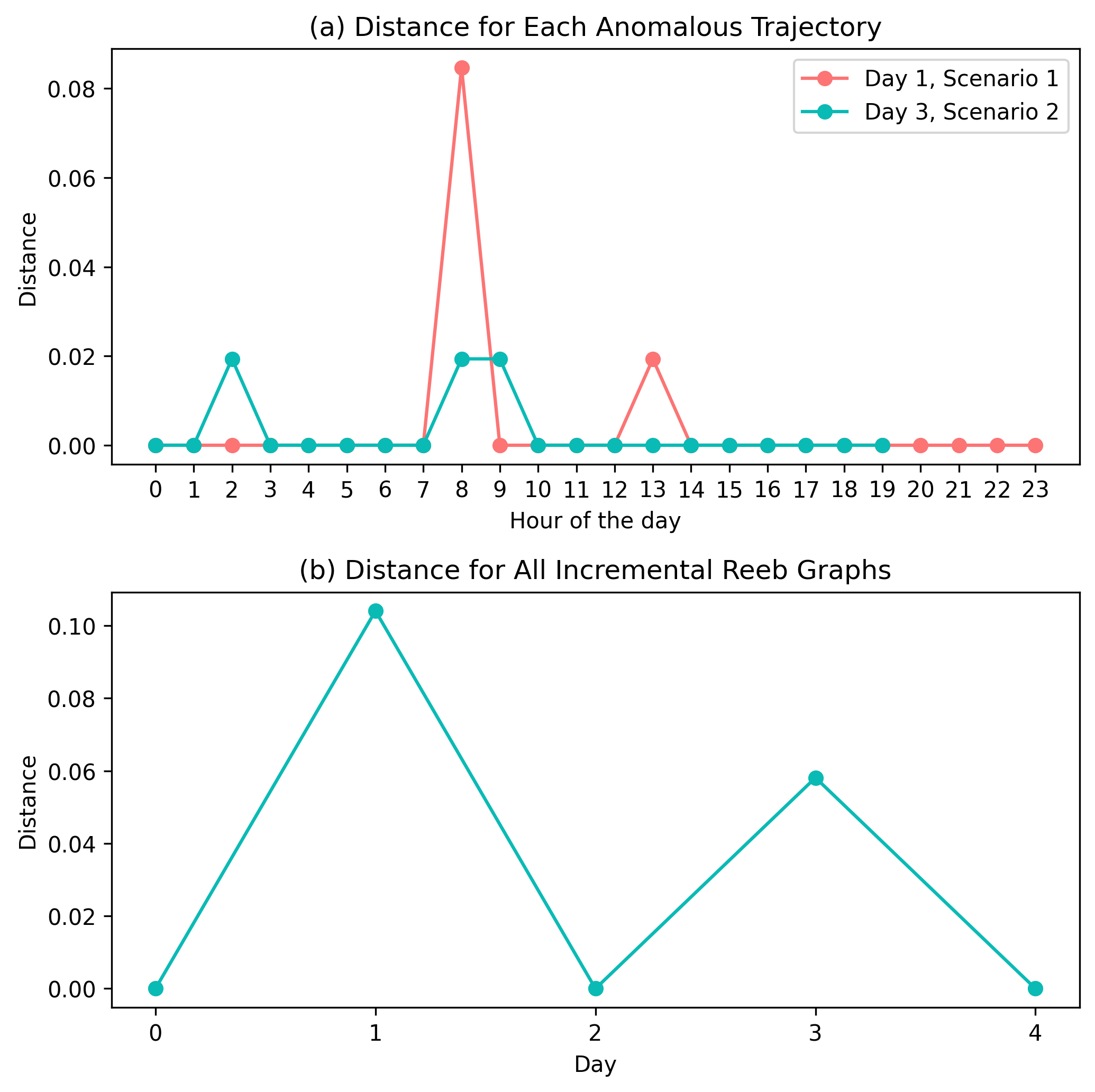}
  \caption{(a) illustrates the Reeb graph node-level distances for both anomalous days.  (b) shows the day-level anomaly scores. }
  \label{fig:incremental}
\end{figure}
\subsection{Quantifying the distance between Reeb graphs}
Given two Reeb graphs, a normal Reeb graph \( R_1 \) and a Reeb graph with one anomalous trajectory \( R_2 \), each containing data points across dimensions of time (0 to 23 hours), the following rules are used to calculate the distance between Reeb graphs defined as $d(R_1, R_2$):

\begin{enumerate}
    \item For each hour, if nodes exist in both \( R_1 \) and \( R_2 \), calculate the Euclidean distance between the nodes.
    \item If only one of the Reeb nodes graphs, \( R_1 \) or \( R_2 \), has a node at a particular hour, calculate the distance to the temporally closest node from the other Reeb graph.
    \item If neither Reeb graph has a node for a given hour, the distance is 0.
\end{enumerate}

Specifically, in point 2 above, we have a case where a node at time \( t_k \) in Reeb graph \( R_1 \) has no corresponding node in \( R_2 \). We find the Euclidean distance to the nodes in \( R_2 \) at $t_{k+1}$. If there are multiple nodes in \( R_2 \) at $t_{k+1}$ or $t_{k-1}$, then we select the one with the minimum distance. $d(R_1, R_2$) is the sum of the distances computed every hour using the above rules.




\subsubsection{Results}
In this case study, we created a synthetic test dataset to investigate both spatial anomalies (Scenario 1, see Figure \ref{fig:scenarios_2d}(b)) and temporal anomalies (Scenario 3, see Figure \ref{fig:scenarios_2d}(d)). The dataset comprises three days of randomly simulated normal behavior and two days of anomalous behavior. Figure \ref{fig:incremental}(a) illustrates the node-level distances for both anomalous days. On Day 1, new anomalous nodes appear at hour 8 (movie theater) and hour 13 (coming back). Anomalous events on Day 3 occur at hours 2, 8, and 9. Figure \ref{fig:incremental}(b) depicts the day-level anomalies; the anomalous distance for Day 1 is higher than for Day 3, reflecting the student's travel to a more distant location on Day 1, whereas, on Day 3, the anomalies involve the same POIs.

\subsection{Scalability with Reeb Graphs}

We successfully applied ReeSPOT to a simulated dataset that is closer to a real-life distribution. This data is an extended version of the data that we described in this paper for proof-of-concept. Here, instead of modeling weekdays of data sampled every hour, we model the patterns over a month sampled at every 15-second interval. This results in $m = 5760$ and $n = 30$. For this dataset, ReeSPOT models the patterns of daily activities for a simulated population of 800,000 agents. Each agent is processed independently, and the Reeb graphs for the entire dataset were constructed within 7.2 hours, parallel processed across 384 CPU cores~(AMD EPYC 9654 @ 3.7 GHz). We also implemented the spatial Reeb graph, ReeBundle as proposed in ~\cite{shailja2023reebundle} but the quadratic time complexity with respect to $m$ made it computationally challenging. More specifically, for $n = 7$ and $m = 5760$, the Reeb graph construction took around 4 minutes for an agent. ReeSPOT is linear with respect to $m$ and thus for the same problem setting it was able to construct Reeb graphs in approximately 12 seconds on one CPU core. This is an important advantage over spatial Reeb graphs which helps us to apply our method on large-scale datasets. Multi-processing across 384 cores enabled us to construct Reeb graphs in less than 8 hours. We also tested ReeSPOT on medium-sized data with 10,000 agents over a period of one week, Reeb Graphs were computed in approximately 5.5 minutes. The above experiments show the applicability of ReeSPOT in modeling agent's data at different resolutions (weekly, monthly, yearly) and also emphasize the scalability of the proposed algorithm.



%% file: sections/discussion.tex
\section{Discussion and Future Work}
\label{sec:discussion}

In this paper, we proposed a Reeb graph-based approach (ReeSPOT) to model the patterns of normalcy using day-to-day human trajectory data. The proposed Reeb graphs abstract large-scale spatio-temporal data into a comprehensible topological construct. We design distinct real-life anomalous scenarios, develop trajectory generation methods, and provide a thorough interpretation of Reeb graph results. The parameters of ReeSPOT can control the granularity of the model according to different applications. On the other hand, ReeSPOT depends on the quality of the trajectory, so false positives can impact the accuracy of the model. One explanation for this is the inherent stochasticity of general human behavior.

Another application is a quantifiable sanity check for raw trajectory data such as teleports. We synthesized such scenarios and observed additional nodes in the Reeb graphs. Our experiment setting in this paper is based on the assumption that each agent is independent and the activities conducted by one agent are not related to the other. However, agents in a given population influence the behavior of each other. Such correlations could serve as additional features to our existing model. ReeSPOT has the flexibility to introduce more parameters and features to robustly support the data abstraction. Geo-foundational features describe the nature of each location the agent visited such as residential, commercial, recreational, etc. Nodes of the Reeb graphs can be labeled with such domain-specific information. Such representation can be used as an input to data-driven methods instead of directly using deep learning methods on raw GPS trajectories.

\section{Acknowledgement}
We would like to thank Kin Gwn Lore for the invaluable insights and assistance throughout this project. This work is supported by the Intelligence Advanced Research Projects Activity (IARPA) via Department of Interior/ Interior Business Center (DOI/IBC) contract number 140D0423C0057 The U.S. Government is authorized to reproduce and distribute reprints for Governmental purposes notwithstanding any copyright annotation thereon. Disclaimer: The views and conclusions contained herein are those of the authors and should not be interpreted as necessarily representing the official policies or endorsements, either expressed or implied, of IARPA, DOI/IBC, or the U.S. Government.